\documentclass[lettersize,journal]{IEEEtran}
\usepackage{amsmath,amsfonts}
\usepackage{color}
\usepackage{array}
\usepackage{algorithmic}
\usepackage[linesnumbered,ruled,vlined]{algorithm2e}
\usepackage[caption=false,font=normalsize,labelfont=sf,textfont=sf]{subfig}
\usepackage{textcomp}
\usepackage{stfloats}
\usepackage{url}
\usepackage{times}
\usepackage{verbatim}
\usepackage{graphicx}
\usepackage{longtable}
\usepackage[utf8]{inputenc}
\usepackage[T1]{fontenc}
\usepackage{lipsum}
\usepackage{multirow}
\usepackage{mathtools}
\begin{document}
\makeatletter
\newcommand{\nosemic}{\renewcommand{\@endalgocfline}{\relax}}
\newcommand{\dosemic}{\renewcommand{\@endalgocfline}{\algocf@endline}}
\newcommand{\pushline}{\Indp}
\newcommand{\popline}{\Indm\dosemic}
\let\oldnl\nl
\newcommand{\nonl}{\renewcommand{\nl}{\let\nl\oldnl}}
\makeatother
\title{\huge Study of Clustering Techniques and Scheduling Algorithms with Fairness for Cell-Free MIMO Networks}

\author{Saeed Mashdour and Rodrigo C. de Lamare 

\thanks{R. de Lamare and S. Mashdour are with PUC-Rio, 
Brazil. R. de Lamare is also with the University of York, UK. The emails are smashdour@gmail.com, delamare@puc-rio.br. This work was supported by Funttel/Finep - Grant No. 01.20.0179.00. }} 


\maketitle

\begin{abstract}
In this work, we propose a clustering technique based on information rates for cell-free massive multiple-input multiple-output (MIMO) networks. Unlike existing clustering approaches that rely on the large scale fading coefficients of the channels and user-centric techniques, we develop an approach that is based on the information rates of cell-free massive MIMO networks. We also devise a resource allocation technique to incorporate the proposed clustering and schedule users with fairness. An analysis of the proposed clustering approach based on information rates is carried out along with an assessment of its benefits for scheduling. Numerical results show that the proposed techniques outperform existing approaches.
\end{abstract}

\begin{IEEEkeywords}
Massive MIMO, cell-free, clustering, AP selection, resource allocation, sum-rate.
\end{IEEEkeywords}\vspace{-0.75em}

\section{Introduction}

The concept of cell-free massive multiple-input multiple-output (CF-mMIMO) networks, initially presented in the works of ~\cite{ngo2015cell, ngo2017cell}, emerged as a promising approach to address requisites such as ubiquitous high data rates, consistent quality of service (QoS), and high reliability. {In CF networks}, numerous access points (APs) serve a smaller number of user equipments (UEs) using identical time-frequency resources. {The CF-mMIMO concept} is suitable for improving the coverage and providing a more uniform performance across UEs. 

{To effectively obtain the benefits of CF-mMIMO networks while ensuring manageable complexity and signaling, the literature describes several clustering strategies \cite{ammar2021user, guevara2021partial, bjornson2020scalable,tentu2022uav, wei2022user,tds, itapsprec&pa, rmmsecf,mashdour2022MMSE-Based, mashdour2022enhanced, mashdour2022multiuser, rscf,cl&sched}. In order to achieve high performance and reduced complexity in comparison to CF networks, the user-centric CF (UCCF) strategy that utilizes clustering based on UEs has been advocated \cite{ammar2021user, guevara2021partial, bjornson2020scalable,tentu2022uav, wei2022user}, ensuring that each UE is supported by a designated cluster of APs. In this strategy, various criteria are applied for AP clustering.} In \cite{ammar2021downlink}, AP clustering is done by defining a specific serving cluster of APs for each UE based on the large scale fading (LSF). {The work of \cite{zhou2022successful} determines AP clustering using the signal to noise ratio of the UE and stochastic geometry theory.}
The work in \cite{banerjee2022access} uses multi-agent reinforcement learning (MARL) for APs to autonomously decide which UEs to serve in dynamic CF-mMIMO networks, showing improved AP clustering performance. 

Another crucial aspect of CF networks is resource allocation to ensure a fair distribution of resources among UEs and optimize spectral efficiency. In~\cite{van2020joint}, the study focuses on resource allocation in CF-mMIMO networks, presenting an algorithm for power reduction that maintains UEs downlink spectral efficiency needs. The work of \cite{wu2021revenue} introduced a resource allocation strategy for network slicing-based CF-mMIMO networks to optimize infrastructure operator revenue while preserving network service quality. In \cite{ammar2021downlink}, resource allocation in CF-mMIMO networks have been addressed by formulating the problem as a weighted sum rate maximization, ensuring efficient utilization of network resources to enhance overall system performance. {In \cite{d2020user}, user association in scalable CF-mMIMO systems was considered for uplink scenarios using the Hungarian algorithm to maximize the system's uplink sum-rate.} In ~\cite{mashdour2022MMSE-Based}, a sequential multiuser scheduling and power allocation (SMSPA) scheme was proposed to enhance the sum-rate performance of the CF and clustered CF MIMO networks. 



{In this work, we employ a user-centric clustering approach for downlink scenarios in CF-mMIMO networks, selecting APs for each UE based on a threshold of information rates, which ensures individualized service quality and efficiently utilizes network resources. Unique to our method is a constraint that increases the minimum number of APs dedicated to serving each UE. This approach ensures a higher service quality, as it guarantees that each UE has sufficient AP coverage.} We also devise a resource allocation technique to incorporate the proposed clustering and schedule users with fairness. An analysis of the proposed clustering approach based on information rates is carried out along with an assessment of its benefits for scheduling. Numerical results show that the proposed techniques outperform existing approaches.

The rest of this paper is organized as follows: In Section \ref{secII}, the system model including different network structures and the related sum-rate expressions are presented. In Section \ref{secIII}, the proposed technique for AP selection is introduced. Section \ref{secIV} presents the resource allocation problem that schedules multiple users with fairness. In Section \ref{secV}, the results are presented and discussed, and Section \ref{secVI} draws the conclusions.
\vspace{-6mm}
\section{System Models} \label{secII}

In this section, we describe CF-mMIMO and UCCF network models along with their sum-rates.
\vspace{-3mm}
\subsection{CF-mMIMO Network Model}

{We describe the downlink of a CF-mMIMO network with $L$ APs arranged randomly, each equipped with $N$ antennas uniformly spaced and $K$ single-antenna UEs randomly distributed. It is assumed that the number of UEs greatly exceeds $M=LN$ as the total number of AP antennas, i.e., $K \gg M$. Thus, it is necessary to schedule $n\leq M$ out of $K$ UEs.}

The channel coefficient linking the $m$th AP {antenna} and the $k$th UE is given by $g_{m,k} = \sqrt{\beta_{m,k}}h_{m,k}$, where $\beta_{m,k}$ is the large-scale fading coefficient, and $h_{m,k} \sim \mathcal{CN}(0, 1)$ is the small-scale fading coefficient defined as independent and identically distributed (i.i.d.) random variables (RVs) that are constant within a coherence interval and independent across different coherence intervals~\cite{ngo2017cell}.  
The downlink signal is expressed as
\vspace{-2mm}
\begin{equation}
{\mathbf{y}=\sqrt{\rho_{f}}\mathbf{G}^T\mathbf{P}\mathbf{x}+\mathbf{w}},
\end{equation}
where $\rho_{f}$ is the maximum transmitted power of each antenna. The channel matrix $\mathbf{G}=\hat{\textbf{G}}+\tilde{\textbf{G}}\in \mathbb{C}^{{LN}\times n}$ encompasses the channel estimate $\hat{\textbf{G}}$ and estimation error $\tilde{\textbf{G}}$ which models CSI imperfections. {The entries of $\hat{\textbf{G}}$ and $\tilde{\textbf{G}}$ are independent zero-mean variables and the estimation error is assumed to remain sufficiently small for communication.} The entries of $\mathbf{G}$ are denoted by $\left [ \mathbf{G} \right ]_{m,k}=g_{mk}$. $\mathbf{P}\in \mathbb{C}^{{LN}\times n}$ is the linear precoder matrix, and $\mathbf{x}=\left [ x_{1},\cdots ,x_{n} \right ]^{T}$ is the zero-mean symbol vector with $\mathbf{x}\sim \mathcal{CN}\left ( \mathbf{0},\mathbf{I}_{n} \right )$ and $\mathbf{w}=\left [ w_{1},\cdots,w_{n} \right ]^{T}$ is the additive noise vector, where $\textbf{w}\sim \mathcal{CN}\left ( 0,\sigma_{w}^{2}\textbf{I}_{n} \right )$. We note that several linear and non linear precoders \cite{mmimo,wence,joham,gbd,wlbd,cqabd,mbthp,rmbthp,siprec,rprec&sr,rsbd,rsthp,rapa,lrcc,1bitprec,zcprec} could be considered in this context. We assume Gaussian signaling, statistical independence among elements of $\mathbf{x}$, as well as independence from all noise and channel coefficients. Consequently, the sum-rate of the CF system is computed as
\begin{equation}\label{eq:RCF}
SR_{cf}=\log_{2}\left ( \det\left [\textbf{R}_{cf}+\textbf{I}_n \right ]\right ),
\end{equation}
\vspace{-2mm}
where the covariance matrix $\textbf{R}_{cf}$ is given by
\begin{equation}\label{eq:RCF_1}
\textbf{R}_{cf}=\rho _{f} \hat{\textbf{G}}^{T}\textbf{P}\textbf{P}^{H}\hat{\textbf{G}}^{\ast }\left ( \rho _{f}\tilde{\textbf{G}}^{T}\textbf{P}\textbf{P}^{H}\tilde{\textbf{G}}^{\ast } +\sigma _{w}^{2}\textbf{I}_n\right )^{-1}.
\end{equation}
\vspace{-8mm}

\subsection{UCCF Network Model}
\begin{figure}
    \centering
    \includegraphics[width=.96\linewidth]{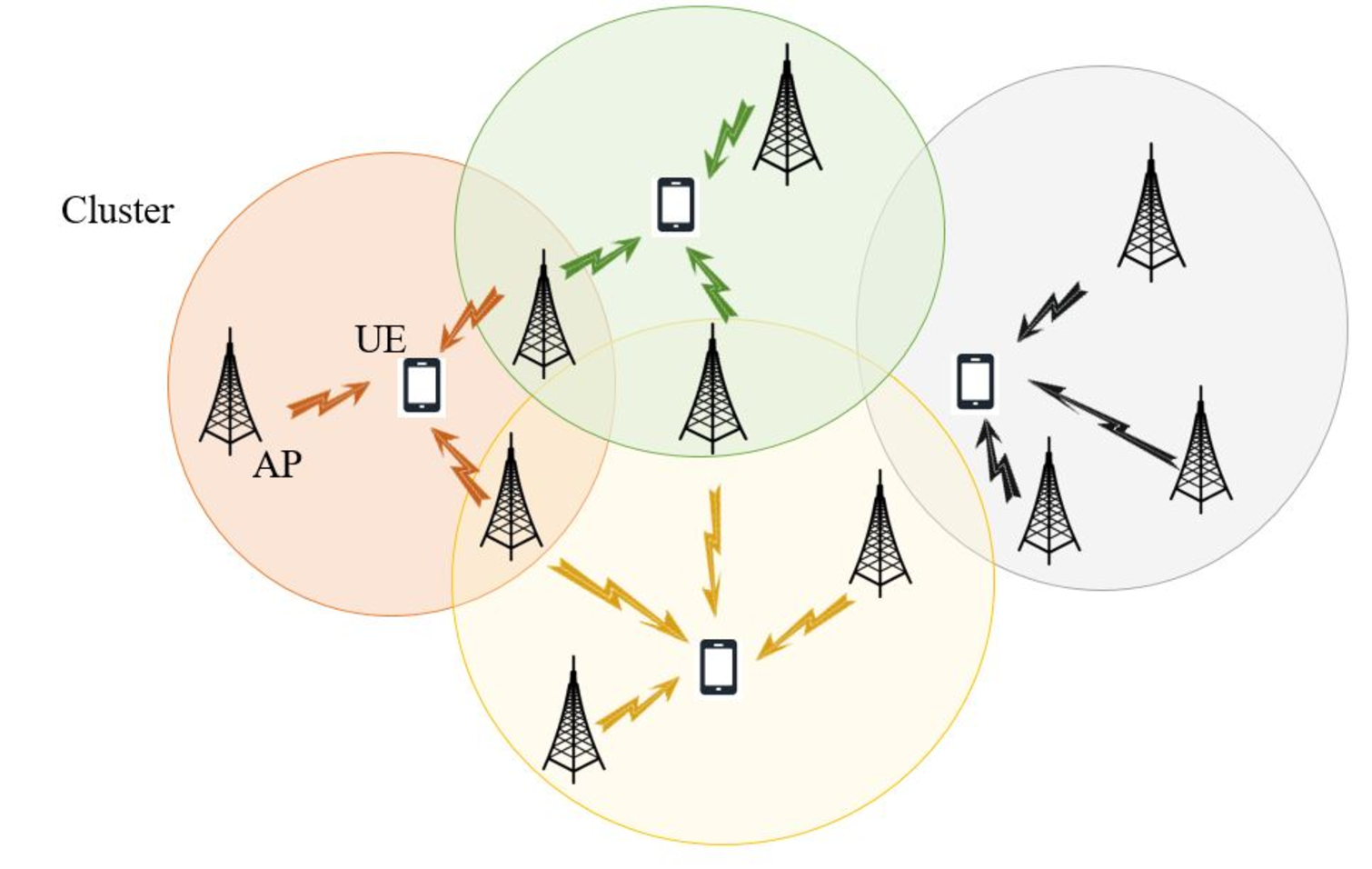}
    \vspace{-1em}
    \caption{User-centric cell free network.}
    \vspace{-1em}
    \label{model}
\end{figure}

{We utilize a UCCF massive MIMO network as illustrated in Fig.~\ref{model}, where each UE is served by a subset of APs. The downlink received signal in this network is modeled as:}
\vspace{-2mm}
\begin{equation} \label{eq:CF-sig}
    \textbf{y}=\sqrt{\rho _{f}}\textbf{G}_{a}^T\textbf{P}_a\textbf{x}+\textbf{w}.
\end{equation}
Similar to the CF network, the sum-rate of the UCCF system is given by
\vspace{-2mm}
\begin{equation}\label{eq:RCF}
    R_{UC}=\log_{2}\left (  \det\left [\textbf{R}_{UC}+\textbf{I}_K  \right ]\right ),
\end{equation}
where the matrix $\textbf{R}_{UC}$ is
\vspace{-2mm}
\begin{equation}\label{eq:RCF_1}
    \textbf{R}_{UC}=\rho _{f} \hat{\textbf{G}}_a^{T}\textbf{P}_{a}\textbf{P}_{a}^{H}\hat{\textbf{G}}_a^{\ast }\left ( \rho _{f}\tilde{\textbf{G}}_a^{T}\textbf{P}_{a}\textbf{P}_{a}^{H}\tilde{\textbf{G}}_a^{\ast } +\sigma _{w}^{2}\textbf{I}_K\right )^{-1},
\end{equation}
and $\textbf{x}$ and $\textbf{w}$ are statistically independent.

{The channel and precoder matrices in a UCCF network are considered as $\textbf{G}_{a}=\left [ \textbf{g}_{a1},\cdots , \textbf{g}_{an}\right ]$ and $\textbf{P}_{a}$ as a function of $\textbf{G}_{a}$,  respectively, where $\textbf{g}_{ak} \in \mathbb{C}^{{LN}\times 1}$, $k \in \left \{ 1,\cdots ,n \right \}$. We define  $\textbf{g}_{ak}=\textbf{A}_{k}\textbf{g}_{k}$ where $\textbf{A}_{k}=\textup{diag}\left ( a_{k1},\cdots ,a_{k{LN}} \right )$ are diagonal matrices with entries given by}
\vspace{-2mm}
\begin{equation} \label{akl}
    a_{kl}=\left\{\begin{matrix}
1 & if \  l\in U_{k}\\ 
0 & if \  l\notin  U_{k}
\end{matrix}\right.
, l\in \left \{ 1,\cdots ,{LN} \right \},
\end{equation}
where $U_{k}$ denotes the subset of APs serving $k$th UE, $\textbf{g}_{k} \in \mathbb{C}^{{LN}\times 1}$, $k \in \left \{ 1,\cdots ,n \right \}$ is the $k$th column of the ${LN}\times n$ CF channel matrix $\textbf{G}=\left [ \textbf{g}_{1},\cdots , \textbf{g}_{n}\right ]$.

\section{Proposed Clustering Based on Information Rates} \label{secIII}

The selection of APs for serving UEs in wireless networks, known as clustering,  can be based on various criteria. In this section, we first review standard clustering based on the Large Scale Fading (LSF) criterion. Then, we present a method based on the sum-rate (SR) criterion denoted Boosted SR (BSR) to obtain enhanced AP clustering.

\vspace{-2mm}
\subsection{Clustering Based on Large Scale Fading}

Clustering based on the LSF approach relies on large-scale propagation characteristics like path loss and shadowing to select the potential APs for the UEs. LSF focuses on the average signal reduction over large distances and does not take into account certain features of the wireless environment.

 For AP clustering using this method, AP $m$ with an average channel gain to user
$k$ above a chosen threshold $\alpha_{lsf}$ is considered so that $\beta_{mk}\ge \alpha_{lsf}$ where $\beta_{mk}$ stands for the large scale fading coefficient. For the users which this condition is not met by any AP, a single AP with the largest average channel gain is selected as the corresponding subset \cite{ammar2021downlink}. Thus, the selected AP set for user $k$ is 
\vspace{-2mm}
\begin{equation} \label{eq:CF-sig}
    U_{k,lsf}=\left\{ m \ : \ \beta_{mk}\ge \alpha_{lsf} \ \right\}\cup \left\{ \underset{m}{\text{argmax} \ \beta_{mk}} \right\}
\end{equation}
where the threshold is set as
\vspace{-2mm}
\begin{equation} \label{eq:alphalsf}
    \alpha_{lsf}=\frac{1}{{LN}K}\sum_{m=1}^{{LN}}\sum_{k=1}^{K}\beta_{mk}
\end{equation}
So that the channels with large gains are considered for each user while channels with small gains are discarded \cite{rscf}. {The use of LSF ensures UEs are linked to APs within a close range, accommodating signal variability due to obstacles or distance, and allowing a flexible network structure.} 
{However, this method does not consider instantaneous channel conditions or specific UE requirements. While it has its merits and provides a reasonable account of the wireless environment, it might not offer the optimal AP choices in all scenarios due to its macroscopic perspective. Since it is not customized to instantaneous channel conditions or to enhancing information rates, it might result in losses of  spectral efficiency.}
\vspace{-2mm}
\subsection{Proposed Clustering Based on the Sum-Rate}

In this section, we introduce AP clustering based on the SR criterion. Then an analysis of the proposed approach is presented. 
This approach focuses on the sum-rate that can be achieved between an AP and UEs while ensuring reliable communication. It accounts for various factors such as the actual channel gain, the channel estimation error and noise, which influence the communication quality and information rates. By using SR as a criterion, the method ensures that UEs are served by APs based on their performance and the rate they can offer. The downlink signal model from AP $m$ to UE $k$ is represented by
\vspace{-2mm}
\begin{equation}
    y_{km}=\sqrt{\rho _{f}}\hat{{g}}_{km}^{T}\textbf{p}_{k}\mathbf{x}+\sqrt{\rho _{f}}\tilde{{g}}_{km}^{T}\textbf{p}_{k}\mathbf{x}+w_{k},
\end{equation}
where $\hat{{g}}_{km}$ represents the true channel gain between AP $m$ to UE $k$, $\tilde{{g}}_{km}$ is the estimation error of the channel gain, $\textbf{p}_{k} \in \mathbb{C}^{{LN}\times 1}$ represents the precoder to UE $k$. Assuming Gaussian signaling, the rate from AP $m$ to UE $k$ is given by
\vspace{-2mm}
\begin{equation}\label{eq:SRkm}
    {SR_{km} = \log_{2}\left (  1+\frac{\sqrt{\rho _{f}}\left | \hat{g}_{km} \right |^{2}\textbf{p}_{k}^{H}\textbf{p}_{k}}{\sqrt{\rho _{f}}\left | \tilde{g}_{km} \right |^{2}\textbf{p}_{k}^{H}\textbf{p}_{k}+\sigma _{w}^{2}}\right )}.
\end{equation}
Hence, for effective AP selection, we adopt the average of the rates across all UEs and APs. To ensure efficient coverage and quality of service, every UE should ideally be served by an AP or a cluster of APs that can offer a rate greater than or equal to an average rate $\alpha _{src}$ defined as
\vspace{-2mm}
\begin{equation} \label{alpha-asr}
    \alpha _{src}= \frac{1}{KM}\sum_{k=1}^{K}\sum_{m=1}^{M}SR_{km}
\end{equation}
Indeed, for each UE, the APs with rates exceeding $\alpha _{src}$ are designated as the AP cluster catering to that UE. However, there might be scenarios that no AP can offer a rate meeting this criterion for a particular UE. To address such situations, the UE will be served by the AP that yields the highest rate even if it is below the desired benchmark. Consequently, the cluster of APs chosen to serve UE $k$ is defined as
\vspace{-2mm}
\begin{equation} \label{ASR}
    U_{k,asr}=\left\{ m \ : \ SR_{km}\ge \alpha _{src} \ \right\}\cup \left\{ \underset{m}{\text{argmax} \ SR_{km}} \right\}.
\end{equation}
In basic terms, this method prioritizes the information rates over traditionally used metrics ensuring that UEs are served by the APs that can provide the best information rates in any scenario. To ensure that the approach in \eqref{ASR} works well, we also impose a constraint that inscreases the minimum number of APs dedicated to serving each UE. {The BSR algorithm uses a dynamic approach to AP clustering, prioritizing sum rates for communication quality and network efficiency.} The BSR algorithm is summarized in Algorithm \ref{alg:algBASR}, where $\mathcal{K}$ and $\mathcal{A}$ show the sets of all UEs and all APs, respectively. 

\begin{algorithm}
\LinesNumbered
\SetKwBlock{Begin}{}{}
\caption{Boosted SR (BSR) Algorithm}
\label{alg:algBASR}
\SetAlgoLined
\nonl 
\Begin{Input rate matrix \textbf{SR}, threshold 
\( \alpha_{src} \)\\
    \textbf{Output:} Allocated AP set for each UE \( AP_{k}^{\text{ASR}} \)\\
    For every UE \( k \), assign an empty cluster \( AP_{k}^{\text{BSR}} \)
    
    \textbf{Step 1: SR calculation and initial AP selection}\\
    
   \textbf{Step 2: Evaluation of AP coverage}\\
    
    \textbf{Step 3: Identification of under-supported UEs}\\
    
   \textbf{Step 4: Augmenting AP coverage}\\
    }
    Calculate sum-rate \\ 
    
\end{algorithm}

\subsection{Analysis of the BSR Algorithm}
 
In AP clustering based on BSR, by selecting APs that provide the highest rates, the system is directly maximizing the network's sum-rate. This method adjusts to actual conditions of the network. This makes BSR based clustering aim for the best possible performance based on the current conditions of the network. It also has the potential to offer better spectral efficiency because BSR considers the information rates.
To analyze the BSR method, we consider eq. (\ref{eq:SRkm}) similar to ${SR_{km} = \log_{2}\left (  1+\frac{\left \|S_{km}  \right \|^{2}}{\left \|IN_{km}  \right \|^{2}}\right )}$, the rate for two distinct links $\left \{ j,i \right \}$ to UE $k$ is
\vspace{-3mm}
\begin{equation} \label{SRkj}
    {SR_{kj} = \log_{2}\left (  1+\frac{\left \|S_{kj}  \right \|^{2}}{\left \|IN_{kj}  \right \|^{2}}\right )}
\end{equation}
\vspace{-3mm}
\begin{equation} \label{SRki}
    {SR_{ki} = \log_{2}\left (  1+\frac{\left \|S_{ki}  \right \|^{2}}{\left \|IN_{ki}  \right \|^{2}}\right )}
\end{equation}
If we assume that 
\begin{itemize}
   \item $\left \| S_{kj} \right \|^{2}$ is slightly larger than $\left \| S_{ki} \right \|^{2}$ (i.e., link $j$ has a slightly better channel)
    \item However, $\left \| IN_{kj} \right \|^{2}$ is significantly larger than $\left \| IN_{ki} \right \|^{2}$, (i.e., link $j$ has a lot more noise)
\end{itemize}
Then, we could have $SR_{ki}>SR_{kj}$, resulting in
\begin{itemize}
    \item LSF would choose link $j$ because $\left \| S_{kj} \right \|^{2}>\left \| S_{ki} \right \|^{2}$.
    \item BSR would choose link $i$ because $SR_{ki}>SR_{kj}$.
\end{itemize}
In scenarios where the noise variance is large, merely selecting an AP based on a large channel gain might not yield the best sum-rate performance. By considering both the channel conditions and the noise levels and consequently using SR, we are more likely to choose the AP that provides a better link for the UE, resulting in better system sum-rate performance.

{We assume that LSFs from all antennas of an AP to a UE are identical. Consequently, 
the set $\left \{ a_{k\left ( j-1 \right )N+1},\cdots , a_{jN} \right \}$ in $\textbf{A}_{k}$ comprises solely ones when AP $j$ is serving UE $k$, and zeros otherwise. In contrast, BSR integrates small scale fading into its evaluation which varies across different antennas of APs, providing an adaptive assessment of each AP's performance. {The BSR algorithm improves network performance by optimizing AP selection for
UEs, ensuring balanced and efficient network load with high-quality connections. By dynamically assigning UEs to APs based on information rates and enhancing coverage for under-supported UEs, BSR not only ensures reliability through adaptive coverage but accounts for the unique fading characteristics of each antenna. In contrast to the static nature of the LSF-based criterion, BSR leads to a network that has higher sum rates and enhanced reliability.}

}

\vspace{-3mm}

\section{Resource Allocation with Fairness} \label{secIV}

In CF-mMIMO networks with $K \gg {LN}$, clustering and multiuser scheduling can play an important role together. Since a key goal of CF-mMIMO networks is to provide more uniform performance across UEs, we develop a multiuser scheduling algorithm to ensure fairness across UEs. However, the order in which the clustering and multiuser scheduling are performed result in different costs. Due to its reduced costs, we perform clustering followed by multiuser scheduling. 
\vspace{-3mm}
\subsection{{Proposed Resource Allocation Problem}}

In each time slot $i$, our goal is to schedule a subset of $n$ UEs from a total of $K$ UEs, where $n \leq {LN}$, to achieve a desirable sum-rate. The selected UEs are represented as $S_n^i$, resulting in a column-reduced channel matrix $\textbf{G}_{cc}(S_n^i)$. The objective is to maximize the sum-rate of the selected UEs in a UCCF network while ensuring fairness as formulated by\vspace{-2mm}
\begin{subequations}
\begin{align}
\underset{S_{n}^i, \textbf{d}}{\text{maximize}} ~ & SR_{UC} (S_{n}^i)\\
\text{subject to} & \left \| \textbf{P}_a \left (S_{L}^i \right ) \right \|_{F}^{2}\leq P, \forall i=1, \cdots, T \label{b}\\
&  
{S}_{n}^i \cap {S}_{n}^j=\emptyset, i \neq j, \forall i,j=1, \cdots, T \label{c}\\
&
\cup_{i=1}^{T} {S}_{n}^i=\mathcal{K} \label{d}\\
&  
\frac{1}{\lvert \mathcal{S}_{c}\rvert}\sum_{k \in \mathcal{S}_{c}}{t_{w}}_{k}=\frac{1}{\lvert \mathcal{S}_{p}\rvert}\sum_{k \in \mathcal{S}_{p}}{t_{w}}_{k} \label{e}
\end{align}
\label{opt prob}
\end{subequations}
where $SR_{UC}$ represents the sum-rate for UEs in the $i$th timeslot, with the signal covariance matrix's upper boundary defined by $\textup{Trace}\left [ \textbf{C}_{\textbf{x}}\right ]\leq P$. We have $T$ as the number of timeslots and $\mathcal{K}=\left \{ 1,2, \cdots , K \right \}$ indicating all the UEs. The sets $\mathcal{S}_{c}$ and $\mathcal{S}_{p}$ correspond to UEs contributing to the max sum-rate and those with poor channels, respectively, and ${t_{w}}_{k}$ is the waiting time for UE $k$. Constraints \eqref{c} and \eqref{d} ensure every UE is selected once, while \eqref{e} ensures similar average waiting times for both sets. The problem's complexity arises from the nature of rates and user scheduling challenges.
\vspace{-7mm}
\subsection{Proposed F-Gr Algorithm}

In order to solve the optimization problem in~\eqref{opt prob}, we devise a fair greedy (F-Gr) multiuser scheduling technique based on the greedy (Gr) user scheduling technique in \cite{dimic2005downlink} to maximize the sum-rate considering the constraint in~\eqref{b}. Unlike the approach in \cite{dimic2005downlink}, the F-Gr technique ensures fairness among UEs.

We develop a strategy in F-Gr to enforce equal waiting time for both classes of UEs ($\mathcal{S}_{c}$ and $\mathcal{S}_{p}$) on average, so that after excluding the previously scheduled UEs, we consider a subsequent timeslot to select the UEs with the least channel gain. This strategy is pursued in an alternating fashion until all UEs are scheduled. 

In particular, we consider $\textbf{SR}\in \mathbb{C}^{K\times {LN}}$ as the rate matrix, where $\textbf{SR}_{k,m}=SR_{km}$ as given by (\ref{eq:SRkm}), \(\mathcal{K}\) as the set of all UEs, and calculate the threshold as in (\ref{alpha-asr}). Using the proposed BSR algorithm, APs are clustered for each UE. For the scheduling phase, spanning \(T\)  timeslots, the operation of the \(i\)th timeslot  varies based on whether \(i\) is even or odd. If \(i\) is odd, the system schedules the best \(n\) UEs set $S_{c}^{\frac{i+1}{2}}$ to maximize the sum-rate. The procedure is such that in an odd $i$, considering the $\mathcal{K}_{i}$ as all the UEs in the current timeslot, a UE \(s_{1}\) is identified as follows in the first iteration $l=1$,\vspace{-2mm}
\begin{equation}
s_{1}=\underset{k\in \mathcal{K}_{i}}{\textup{argmax}} 
~ \mathbf{g}_{k}^{H}\mathbf{g}_{k}.
\end{equation}
Denoting $S_{1}=\left \{ s_{1} \right \}$, the achieved rate \(SR\left ( S_{1} \right )_{max}\) is computed and then in the next iteration we find a UE that maximizes the rate as \vspace{-2mm}
\begin{equation}
s_{l}=\underset{k\in\left (  \mathcal{K}_{i} \setminus  S_{l-1} \right )}{\textup{argmax}}SR\left ( S_{l-1}\cup \left \{ k \right \} \right )
\end{equation}
This procedure continues until $n$ UEs are scheduled as the best \(n\) UEs in an odd timeslot. However, in every even timeslot, \(n\) UEs with the poorest channel powers are scheduled as $S_{p}^{\frac{i}{2}}$. After scheduling, the weighted sum-rate of each set is computed and the average sum-rate over all timeslots is \vspace{-2mm}
\begin{equation} \label{SR_av}
    SR_{Av} = \frac{1}{T} \sum_{i=1}^{T} \frac{n_{i}}{n} \times SR^{i}
\end{equation}
The proposed resource allocation framework including clustering and multi-user scheduling is summarized as Algorithm~\ref{alg:cap}. {The greedy selection in the F-Gr algorithm, efficiently balances computational complexity with high performance for scalable use in large networks and ensures convergence via a finite set of UEs and deterministic scheduling steps.} \vspace{-0.5em}

\begin{algorithm}
\LinesNumbered
\SetKwBlock{Begin}{}{}
\caption{Proposed {F-Gr} Resource Allocation}\label{alg:cap}
\SetAlgoLined
\nonl 
\vspace{-1em}
\Begin{Input rate matrix $\textbf{SR}$, Set \( \mathcal{K}_{1} = \mathcal{K} \)\\
    Calculate threshold using eq. (\ref{alpha-asr}) \\
    Cluster APs per UE using BSR \\
    Determine \(n\) (UEs scheduled per timeslot) \\

    \For{\(i = 1\) to \(T\)}{
        \If{\(i \mod 2 = 1\)}{
            \( l = 1, s_{1} = \underset{k \in \mathcal{K}_{i}}{\textup{argmax}} \mathbf{g}_{k}^{H} \mathbf{g}_{k}, S_{1} = \{ s_{1} \} \)\\
            \While{\( l < n \)}{
                \( l = l + 1 \) \\
                \( s_{l} = \underset{k \in ( \mathcal{K}_{i} \setminus S_{l-1} )}{\textup{argmax}} SR( S_{l-1} \cup \{ k \} ) \) \\
                Update \( S_{l} = S_{l-1} \cup \{ s_{l} \} \) \\
                
                \If{\( SR( S_{l} ) \leq SR( S_{l-1} ) \)}{
                    Break, \( l = l - 1 \)
                }
            }
            \( S_{c}^{\frac{i+1}{2}} =  S_{l} \) \\
            Update UEs: \( \mathcal{K}_{i} = \mathcal{K}_{i} \setminus S_{c}^{\frac{i+1}{2}} \)
        }
        \Else{
            Schedule \( n \) UEs with poorest channels \( S_{p}^{\frac{i}{2}} \) \\
            Update UEs: \( \mathcal{K}_{i} = \mathcal{K}_{i} \setminus S_{p}^{\frac{i}{2}} \)
        }
    }

    Calculate sum-rate for scheduled UEs \\
    Compute average sum-rate using eq. (\ref{SR_av}) \\
}

\end{algorithm} 

\section{Simulation Results} \label{secV}
\vspace{-1mm}

\begin{figure}
	\centering
\includegraphics[width=.96\linewidth]{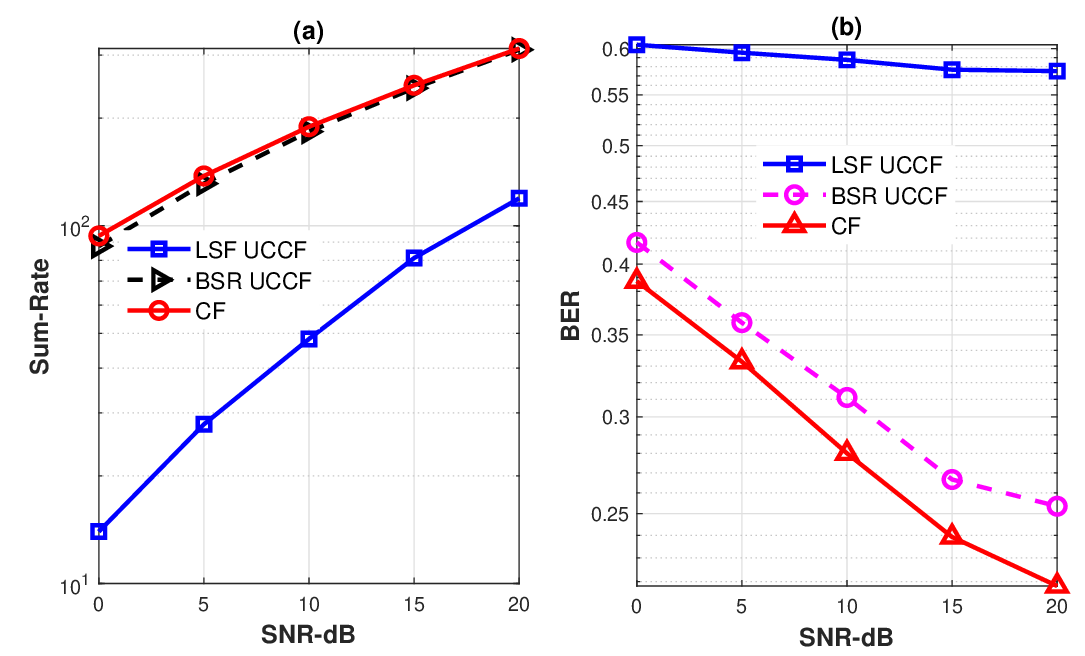}
 \vspace{-3mm}
	\caption{{{Performance of networks: (a) Sum-rate for UCCF with different clustering criteria and CF network with no user scheduling, $L=16$, $N=4$, $K=128$}, (b): BER of UCCF networks with different clustering criteria and the CF network with no user scheduling, $L=16$, $N=4$, $K=128$.}}
	\label{fig:LSF-ASR-BASR} 
\end{figure}

In this section, we assess the sum-rate performance of the clustering techniques. {We consider a CF network in a square area with side length of 400~m, $L=16$ APs each equipped with $N=4$ antennas, imperfect CSI and $K=128$ randomly distributed single-antenna UEs.} {For the analysis of both sum-rate and BER performances, our results are averaged over 1000 channel realizations, ensuring the statistical reliability and robustness of the results.} 
Fig.~\ref{fig:LSF-ASR-BASR} (a) depicts a comparison of UCCF networks using the LSF and BSR criteria with the CF network where no user scheduling is performed. BSR greatly outperforms LSF and approaches the CF network. {For the same network without scheduling, Fig.~\ref{fig:LSF-ASR-BASR} (b) presents the bit error rate (BER) curves, illustrating a substantial BER improvement of BSR over LSF.}

\begin{figure}
	\centering
\includegraphics[width=.96\linewidth]{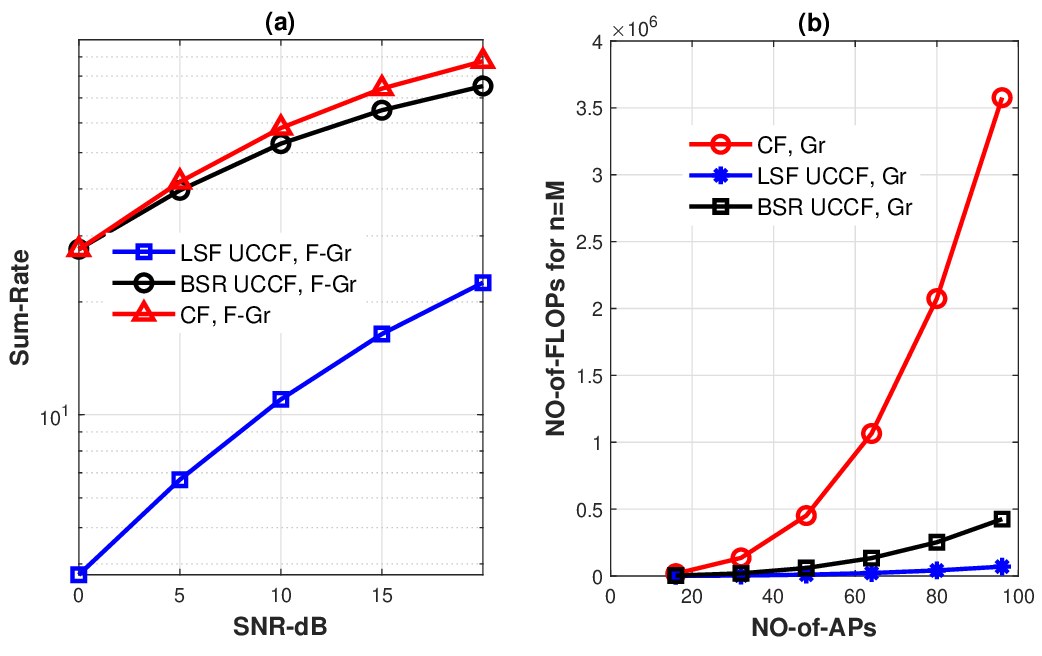}
\vspace{-1em}
	\caption{Performance and complexity of F-Gr resource allocation: (a) F-Gr resource allocation sum-rate in CF and UCCF networks, $L=16$, $N=4$, $K=128$ and $n=20$, (b) Complexity of the proposed resource allocation for CF and UCCF networks when $n=LN$ UEs are scheduled. }
	\label{fig:F-Gr} 
\end{figure}

We evaluated the F-Gr resource allocation algorithm by examining its performance in both CF and UCCF networks, the latter utilizing BSR and LSF criteria for AP clustering. Accordingly, Fig.~\ref{fig:F-Gr} (a) shows the performance comparison of F-Gr in CF, the UCCF network with BSR clustering and the UCCF network with LSF clustering for configurations $M=64$ APs and $K=128$ UEs, and scheduling $n=20$ UEs per timeslot. It is shown that using the proposed F-Gr algorithm notably outperforms the LSF UCCF network and approaches the performance of the CF network. In Fig.~\ref{fig:F-Gr} (b) the computational cost in terms of flops is shown for CF, BSR UCCF and LSF UCCF networks for scheduling $n=LN$ UEs per network. It shows that UCCF network has a much less complexity compared to the CF network, {and reveals the scalability and efficiency of BSR and F-Gr, even with up to 100 APs, ensuring high performance in a dense network.} It also shows that LSF with UCCF networks is less complex than BSR with UCCF networks. This is because BSR with UCCF networks uses both large scale and small scale fading information, which results in significant performance improvement. Accordingly, Table.~\ref{table:complexity} shows the calculated computational cost in terms of flops and the signaling load for these networks which shows that using LSF results in less signaling load compared to BSR.

\vspace{-1mm}
\begin{table}[htb!]
\vspace{-1em}
 \caption{ Computational complexity and signaling load of the proposed resource allocation algorithm}
\vspace{-2.5em}
\begin{center}
{\scriptsize 
\begin{tabular}{| m{4.5em} | m{5cm} | m{6em} |} 

\hline
Network & NO of FLOPs & Signaling Load \\
\hline
\hline

CF, {Gr} & $4(LN)^{3}+LN\left ( 2K+6 \right )$ & \\
\hline

LSF UCCF, {Gr} & $\frac{9}{128}(LN)^{3}+LN\left ( \frac{3}{8}K+7 \right )+K+1$ & $2N^{2}L^{2}+NL^{2}+NL$ \\
\hline

BSR UCCF, {Gr} & $\frac{27}{64}(LN)^{3} + \frac{27}{32}(LN)^{2} + \frac{1179}{256}LN + \frac{19}{8}LNK$ & $4N^{2}L^{2}+2NL$ \\
\hline

\end{tabular}
} 
\label{table:complexity}
\end{center}
\end{table}

\begin{figure}
	\centering	\includegraphics[width=.96\linewidth]{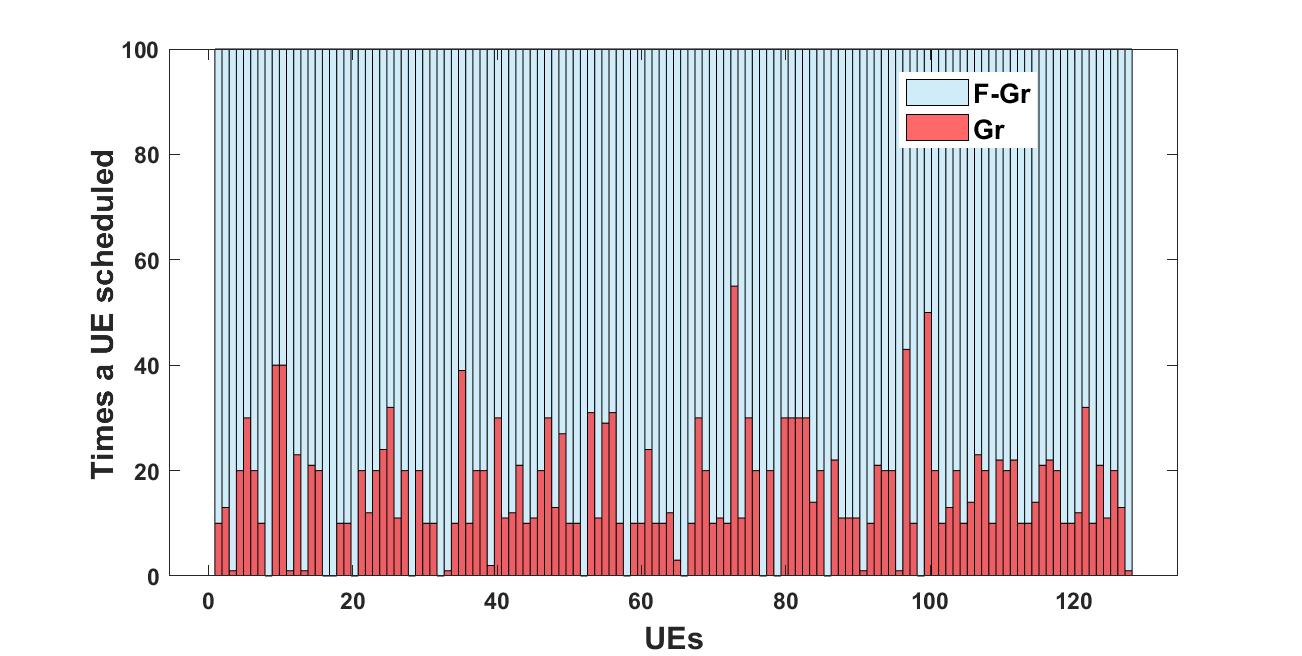} \vspace{-4mm}
	\caption{{{ Scheduling times for each UE in CF network using Gr and F-Gr scheduling methods for $K=128$, $n=20$ and 100 channel realizations.}}} \vspace{-5mm}
	\label{fig:histogram} 
\end{figure}

Fig.~\ref{fig:histogram} considers 100 channel realizations and the number of scheduling times of each user in the CF network. In the standard Gr technique, there are some users which are not supported well in several channel realizations while in the F-Gr technique, all the users are supported in different time slots. \vspace{-1.4em}

\section{Conclusion} \label{secVI}

In this study, we have presented BSR clustering based on information rates and the F-Gr multiuser scheduling algorithm with fairness for UCCF networks. The results reveal that BSR obtains up to $35 \%$ higher rates than the existing LSF, and that F-Gr can ensure fairness across UEs.  \vspace{-4mm}   
\bibliographystyle{IEEEbib}
\bibliography{Refs}

\end{document}